\documentclass[pra,twocolumn,showpacs,preprintnumbers,amsmath,amssymb]{revtex4}

\usepackage{graphicx}
\usepackage{dcolumn}
\usepackage{bm}

\def \ed {\end{document}}
\def\Fbox#1{\vskip1ex\hbox to 8.5cm{\hfil\fboxsep0.3cm\fbox{%
  \parbox{8.0cm}{#1}}\hfil}\vskip1ex\noindent}  

\def\be{\begin{equation}}\def\ee{\end{equation}}
\def\bea{\begin{eqnarray}}\def\eea{\end{eqnarray}}
\def\bse{\begin{subequations}}\def\ese{\end{subequations}}
\newcommand{\BE}[1]{\begin{equation}\label{#1}}
\newcommand{\BEA}[1]{\begin{eqnarray}\label{#1}}
\newcommand{\BSE}[1]{\begin{subequations}\label{#1}}

\let \= \equiv \let\*\cdot \let\~\widetilde \let\^\widehat \let\-\overline

\def\<{\left\langle}    \def\>{\right\rangle}
\def\({\left(}          \def\){\right)}
 \def \[ {\left [} \def \] {\right ]}

\begin{document}

\title{Nonlinear Dynamics in a Trapped Atomic Bose--Einstein Condensate Induced by an Oscillating Gaussian Potential}

\author{Kazuya Fujimoto}
\affiliation{Department of Physics, Osaka City University, Sumiyoshi-ku, Osaka 558-8585, Japan}%

\author{Makoto Tsubota}
\affiliation{Department of Physics, Osaka City University, Sumiyoshi-ku, Osaka 558-8585, Japan}%
\date{\today}

\begin{abstract}
We consider a trapped atomic Bose--Einstein condensate penetrated by a repulsive Gaussian potential and theoretically investigate the dynamics 
induced by oscillating the Gaussian potential. Our study is based on the numerical calculation of the two-dimensional Gross--Pitaevskii equation. 
Our calculation reveals the dependence of the characteristic behavior of the condensate on the amplitude and frequency of the oscillating 
potential. These dynamics are deeply related to the nucleation 
and dynamics of quantized vortices and solitons. 
When the potential oscillates with a large amplitude, it nucleates many vortex pairs that move away from the potential. 
When the amplitude of the oscillation is small, it nucleates solitons through annihilation of vortex pairs. 
We discuss three issues concerning the nucleation of vortices. The first is the phase diagram for the nucleation of vortices and solitons near the oscillating potential. 
The second is the mechanism and critical velocity of the nucleation. 
The critical velocity of the nucleation is an important issue in quantum fluids, and we propose a new expression for the velocity containing both the coherence length and the size of the potential. 
The third is the divergence of the nucleation time, which is the time it takes for the potential to nucleate vortices, near the critical parameters for vortex nucleation. 

\end{abstract}


\maketitle

\section{INTRODUCTION}
The Gross--Pitaevskii (GP) equation has a long history in the study of quantized vortices and solitons, and is important for both condensed matter and 
nonlinear physics. In superfluid $^3$He and $^4$He, vortices and solitons appear and numerous studies have been made of them \cite{Donnelly,Vollhardt}. However, superfluid $^4$He, which exhibits strong interactions between particles, cannot be quantitatively described by the GP equation since it can be applied only to dilute Bose systems. 

A dilute atomic Bose--Einstein condensate (BEC) has been experimentally realized \cite{Cornell, Ketterle, Hulet} and is quantitatively described by the GP equation. 
Therefore, the dynamics of vortices and solitons expected by the GP equation can be experimentally observed. 
Thus, the experimental realization of dilute atomic BECs sets the stage for the theoretical study of the GP equation. 

Vortices in atomic BECs have been experimentally and theoretically studied. 
The nucleation and dynamics of vortices are mainly treated in two systems. One is a system where the trapping potential rotates and in the other a localized potential moves in the atomic BEC. In the former case, the vortices are nucleated from the surface of the condensate, and the vortex lattice is formed through the characteristic nonlinear dynamics\cite{Tsubota02,Kasamatsu03}. 
In the latter case, vortex pairs are nucleated by the uniform moving potential \cite{Frisch1992,Jackson1998}, giving rise to 
interesting dynamics. Neely $et$ $al$. \cite{Neely10} experimentally and numerically observed that the vortices nucleated by 
the potential migrate with long periodicity in the condensate. The Karman vortex street in atomic BECs has been studied by numerical calculations based on the GP equation \cite{Sasaki10}. 

Solitons have also been investigated in atomic BECs \cite{Anderson01,Huang23,tsuchiya}. Solitons of the GP equation in a one-dimensional system are stable, while those in two or three-dimensional 
systems are unstable against long wavelength transverse oscillation, referred to as sneak instability\cite{tsuchiya, Za}. The instability of solitons leads to the nucleation of vortices, which has been experimentally and numerically confirmed \cite{Anderson01,Huang23}. Thus, solitons are profoundly related to vortices. 

The solitary wave solution of the GP equation has been investigated in detail by Jones and Roberts \cite{Robert1982}. They calculated the energy $E_s$ and the momentum $P_s$ of the 
solution, and found that there are two branches which are drawn from a common point in the $E_{s}$--$P_{s}$ plane, the upper and lower branches, which denote the rarefaction pulse and the vortex 
pair, respectively \cite{Robert1982,Natalia02}. 
We expect that a vortex pair can dynamically change into a rarefaction pulse and vice versa. 
We have numerically observed that annihilation of a vortex pair creates a rarefaction pulse and the collapse of a pulse leads to the nucleation of a vortex pair, which is caused by an oscillating potential \cite{s}. 

A currently important theme in quantum hydrodynamics is quantum turbulence (QT), which was first investigated in superfluid helium \cite{PLTP}. There are a number of methods to create QT in this system. One is to employ thermal counterflow. Another is to oscillate objects such as grids, micro spheres, wires, and forks 
in superfluid helium. Recently, QT in an atomic BEC has been vigorously studied \cite{Berloff02,Kobayashi07,Henn09,Takeuchi10}. There are also several methods of creating QT in this system, namely phase printing \cite{Berloff02}, precessing a trapping potential \cite{Kobayashi07}, and oscillating the potential \cite{Henn09}. 
To date, experimental creation and observation of QT in atomic BECs has been achieved only by oscillating a trapping potential \cite{Henn09}. 

We apply a method for dealing with oscillating objects in superfluid helium to atomic BECs. 
However, there is a definite difference between atomic BECs and helium systems. In helium, moving objects do not nucleate vortices but simply amplify remanent vortices which already exist in the system, creating QT \cite{fujiyama}. 
It is difficult to theoretically treat the nucleation of vortices in helium due to the strong correlation between the atoms and it is not easy to control the event experimentally \cite{Donnelly}. 
In contrast, moving objects intrinsically nucleate vortices in experiments with atomic BECs and optical techniques enable us to observe them \cite{Neely10}.
The nucleation of vortices and their dynamics are quantitatively described by the GP equation.
In this system, a blue detuned laser is used as the object, which is represented by a repulsive Gaussian potential in our numerical calculation. 
We are interested in the nucleation and dynamics of vortices and solitons induced by oscillations of the potential, which should depend on the amplitude and frequency of the oscillating potential.
Previously, we have reported that the oscillating potential causes synergy dynamics of vortices and solitons peculiar to the oscillation, though 
the dependence on the parameters has not been investigated \cite{s}. 
In this paper, through the investigation of the dynamics induced by the oscillating potential, we study the dependence of the dynamics on the parameters and also
the mechanism for the nucleation of vortices, the critical velocity of the nucleation, and the divergence of the nucleation time, which have not been satisfactorily treated for atomic BECs. 

Previous research has been made on the response of atomic BECs induced by an oscillating potential \cite{Raman99, Onofrio00, Jackson00}. 
However, the previous studies focused on dissipation and drag forces and barely considered the instability and nonlinear dynamics induced by an oscillating potential, including dynamical instability, Landau instability, nucleation and dynamics of vortices and solitons, and quantum turbulence. Therefore, our study is very different from previous studies. 

The nucleation of a vortex pair by an oscillating potential can be related to dynamical critical phenomena. In fact, nucleation by a linear moving potential has been investigated in terms of the phenomenon, and a power law between the vortex emission frequency and the velocity of the object near the critical velocity has been reported through a numerical simulation of the GP equation \cite{Huepe00}. 
Thus, we address the nucleation by an oscillating potential in terms of dynamical critical phenomena. Moreover, turbulence is deeply related to pattern formation \cite{pattern}, so that the dynamics induced by the oscillating potential is clearly connected with the process. These phenomena have been historically investigated by the complex Ginzburg--Landau equation, whose typical example is the GP equation. Thus, it is possible to study nonlinear and nonequilibrium phenomena, namely pattern formation and dynamical critical phenomena, in atomic BECs. 
Therefore, our study can pioneer a new direction in atomic BECs. 

The nucleation of vortices in quantum fluids has been considered to be an important issue, including the mechanism and critical velocity of the nucleation. 
Previously, nucleation has been investigated in superfluid $^4$He, where the core size of the vortices is so small that their visualization 
is hard. The theoretical study of nucleation is also difficult since the interaction between the $^4$He atoms is very strong. 
However, in atomic BECs, this situation changes. Optical techniques make visualization possible and the dynamics of vortices can be followed in experiments. 
The nucleation and dynamics of vortices are easily treated because atomic BECs are well described by the GP equation. 
In this paper, we focus on the mechanism and critical velocity of nucleation. The mechanism is discussed in terms of the divergence of the quantum pressure. 
Our numerical calculations reveal that the coherence length and size of the potential are very relevant to the critical velocity 
of the nucleation, and we propose an expression of the velocity including both lengths. 

We consider an atomic BEC confined by a harmonic trap to investigate the dynamics of the condensate induced by a repulsive 
oscillating Gaussian potential. Our study assumes that the condensate is pancake shaped and the temperature of the system is nearly 
zero, which means that the condensate is well described by the two-dimensional GP equation. 

This paper is organized as follows. In Sec. II, we describe the formulation. Section III treats the breakdown of Kelvin's theorem on circulation, which leads to the nucleation of vortices. 
In Sec. IV, the synergy dynamics of vortices and solitons are described. 
The nucleation of solitons near the oscillating potential is discussed in Sec. V. 
Section VI describes the nucleation of multiple vortex pairs. 
A phase diagram summarizing Secs. IV--VI is presented in Sec. VII. 
The critical velocity for the nucleation is treated in Sec. VIII. In Sec. IX, we show the results on divergence of the nucleation time. 
Section X discusses the heating of the condensate. The dependence of the aspect ratio of the condensate on the dynamics, the noise of the system, and 
the appropriate parameters for the growth of QT are discussed in Sec. XI. We summarize our study in Sec. XII. 

\section{FORMULATION}
We consider an atomic BEC near zero temperature, in which the interaction between particles is very weak because of the low 
density and is replaced by an effective interaction proportional to the delta function and the s-wave scattering length. 
Almost all particles are condensed near zero temperature so that this 
system is well described by a macroscopic wavefunction $\psi$ obeying the GP equation.
In this work, we assume that the condensate is strongly confined along the $z$ 
direction, that is, a quasi two-dimensional system. Therefore it is well described by the two-dimensional GP equation: 
\begin{equation}
i\hbar\frac{\partial \psi}{\partial t}
=-\frac{\hbar^2}{2m}\nabla^2 \psi+V\psi
+g \vert \psi \vert^2 \psi ,
\end{equation}
where $m$ is the particle mass, $V$ is the potential, and $g$ is an interaction parameter for the two-dimensional case, where the wavefunction along the 
$z$ direction is assumed to be Gaussian \cite{Kasamatsu03}. The wavefunction 
$\psi$ is normalized by the total particle number $N$. 
We suppose that the condensate is confined by a harmonic potential 
$V_{\rm h}$ and penetrated by a Gaussian potential $V_{\rm G}$, so that $V=V_{\rm h}+V_{\rm G}$ where $V_{\rm h}=\frac{1}{2}
m(\omega _{x}^2x^2+\omega _{y}^2y^2)$ and 
\begin{equation}
V_{\rm G} = V_0{\rm exp}[-\{(x-x_0(t))^2+y^2\}/d^2]. 
\end{equation}
Here $x_{0}(t)$ is the $x$-coordinate of the center of the Gaussian potential and $d$ is its radius. We set the Gaussian 
potential to oscillate as $x_{0}(t)=\epsilon \hspace{0.5mm} \rm{sin}(\omega \it{t})$. We define the velocity of the oscillation as $v = \epsilon \hspace{0.5mm} \omega$. 

We set $ g=4.05 \times 10^{-45} \hspace{0.5mm}\rm{J/m^2} $, $ m=1.42 \times 10^{-25} \hspace{0.5mm}\rm{kg} $, $ N=8.0 \times 10^{4} $, $ \omega _x= 2 \pi \times 4 $\hspace{0.5mm}/s, $ \omega _y= 2 \pi \times 20 $\hspace{0.5mm}/s, $d=1.30 \hspace{0.5mm}\mu \rm m$, and $ V_0= 10gn_0 $ as parameters. Here $n_0$ is the density near the center of the condensate. 
We use a dimensionless form of Eq. (1) to perform a numerical calculation with the Crank--Nicholson method. Space and time are normalized by $\hbar/ \sqrt{2mgn_0}$ and $\hbar/gn_0$, and the space in the $x$ and $y$ directions is discretized into 2048$\times$640 bins. 

\begin{figure}[t]
\begin{center}
\includegraphics[keepaspectratio, width=8cm,clip]{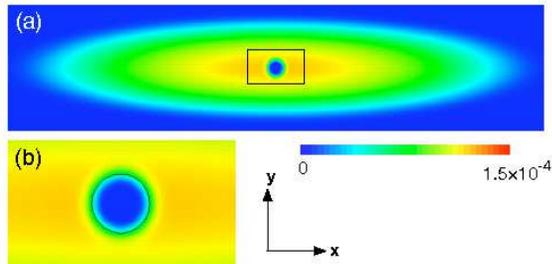}
\end{center}
\caption{(Color online) Initial density profile is shown in (a), where the $x$ and $y$ dimensions of the images are 145 $\mu$m and 34.0 $\mu$m, respectively. 
(b) shows the enlarged figure of the square box in (a), with $x$ and $y$ dimensions of 18.0 $\mu$m and 10.0 $\mu$m, respectively. 
The Thomas--Fermi radii $R_x$ and $R_y$ are 69.3 $\mu$m and 13.9 $\mu$m, respectively. The low density region in the center of the condensate is formed by 
a repulsive Gaussian potential. The diameter of the circle in (b) is 5 $\mu$m.}
\end{figure}

Figure 1 shows the initial density profile in which the low density region of the center is formed by the repulsive potential. 
This is the ground state before the potential starts to move, obtained by the imaginary time step method. The size of the potential $L$ is defined as the rounded value of the diameter of the boundary of the low density region, which is lower than half of the bulk density. In our case, $L = 5\hspace{0.5mm}\mu \rm m$. 
Figure 1(b) shows that the potential of size $L$ is surrounded with a boundary layer, which is a low density region with a width of the order $\xi$. 
This boundary layer plays a key role for any dynamics of vortices near the potential, as discussed later. 

The accuracy of the numerical calculation is confirmed by the two following methods. 
One is the mirror symmetry about the $x$ axis, which must be maintained because the initial state has mirror symmetry and the potential oscillates in the $x$-direction 
to maintain the symmetry. The other is to check that the final state of the condensate, which is obtained by the numerical calculation with forward time evolution, returns 
properly to the initial state in Fig. 1 when we perform the calculation with backward time evolution from the final state. 
This is confirmed by the total energy and the momentum in the $x$ direction.

\section{BREAKDOWN OF KELVIN'S THEOREM ON CIRCULATION}
Vortices can be nucleated in our system because of the breakdown of Kelvin's theorem on circulation. 
This theorem is established in classical fluid dynamics described by the Euler equation and ensures the conservation of vorticity, which leads to neither nucleation nor the disappearance 
of vortices. The breakdown is caused by the divergence of the quantum pressure of the GP equation as shown in the following. 

\begin{figure}[!b]
\begin{center}
\includegraphics[keepaspectratio, width=8cm,clip]{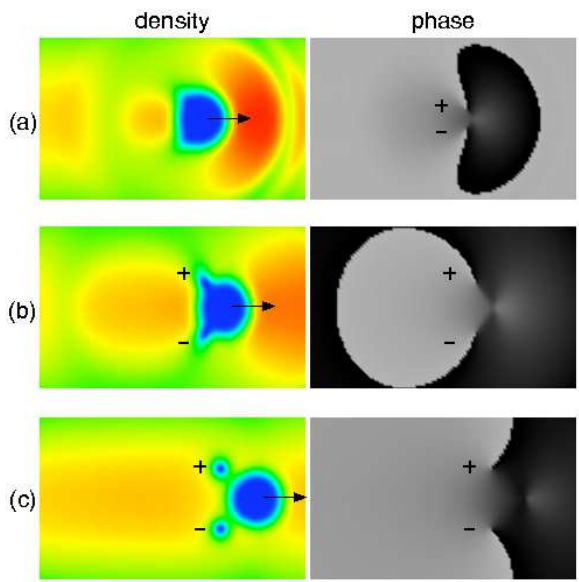}
\caption{(Color online) Nucleation of a vortex pair by the oscillating potential: The density (left) and phase (right) profiles at (a) $t$ = 4.71 ms, (b) $t$ = 8.05 ms , and 
(c) $t$ = 13.7 ms are shown, where the $x$ and $y$ dimensions of each image are 23.0 $\mu$m and 14.0 $\mu$m. The symbols $-$  and $+$ denote a vortex with clockwise or counterclockwise 
circulation, respectively. The black arrows indicate the direction of motion of the potential. The value of the phase varies from $-\pi$ (white) to $\pi$ (black). A ghost vortex pair nucleates inside the potential (a), exits it (b), and finally fully leaves the potential  (c). The parameters of the oscillation are $\epsilon = 10 \mu$m and $\omega = 60 /s$, which are common through Figs. 2--8.}
\end{center}
\end{figure}

First, we derive the general expression for the Lagrange derivative of the circulation in the quantum hydrodynamics described by the GP equation. 
Using the Madelung transformation $\psi = \sqrt{n} \hspace{0.5mm} {\rm{exp}} (i \phi)$ where $n$ and $\phi$ 
are the density and the phase of the wavefunction, we obtain from Eq. (1) the Euler-like equation
\begin{equation}
m \frac{\partial}{\partial t} \bm{v} = - \bm{\nabla} ( \tilde{\mu} + \frac{1}{2} m \bm{v} ^2 )
\end{equation}
with
\begin{equation}
\tilde{\mu} = V + n g - \frac{\hbar ^2}{2 m \sqrt{n}} \bm{\nabla} ^2 \sqrt{n}, 
\end{equation}
where the superfluid velocity $\bm{v}$ is given by ${\hbar} \bm{\nabla} \phi /m$. 
The characteristic term of the equation is the quantum pressure term $\frac{\hbar ^2}{2 m \sqrt{n}} \bm{\nabla} ^2 \sqrt{n}$, which does not appear in the Euler equation. 
Using these equations and the circulation $\kappa (C(t)) = \int _{C(t)} \bm{v} \cdot d\bm{l}$, we can calculate the Lagrange derivative of the circulation 
\begin{equation}
\frac{D}{Dt} \kappa(C(t)) = \lim _{\delta t \to 0} \frac{1}{\delta t} [\int _{C(t+\delta t)} \bm{v}(\bm{r},t+\delta t) \cdot d\bm{l} - \int _{C(t)} \bm{v}(\bm{r},t) \cdot d\bm{l}], 
\end{equation}
where $C(t)$ is a closed contour, which rides the superfluid velocity field to change the configuration, defined by $\bm{r} = \bm{p}(s,t)$ and $\partial \bm{p}(s,t) / \partial t = \bm{v}(\bm{p}(s,t),t)$. Here $s$ is a time-independent parameter denoting a point on $C(t)$, and its change through the domain $s_{0} \leq s \leq s_{1}$ with arbitrary constants $s_{0}$ and $s_{1}$ represents the whole closed contour $C(t)$ \cite{Ke}. 
We rewrite Eq. (5) with the parameter $s$ to transform the integration range of the two terms to the same range of the domain 
$s_{0} \leq s \leq s_{1}$, which make possible the subtraction in Eq. (5). 
As a result, we obtain the following expression: 
\begin{equation} 
\frac{D}{Dt} \kappa(C) = - \frac{1}{m} \int _{C}  \bm{\nabla} ( \tilde{\mu} ) \cdot  d\bm{l} + \int _{C} \{ (\bm{\nabla} \times \bm{v}) \times \bm{v} \} \cdot d\bm{l} + \int _{C} \bm{v} \cdot d\bm{v}. 
\end{equation} 
This is the general expression for the Lagrange derivative of the circulation in the quantum hydrodynamics described by the GP equation.

Second we consider the condition of the nucleation of vortices with Eq. (6), which is derived by the breakdown of Kelvin's theorem on circulation. 
Now, we treat the system without vortices to consider the condition of the nucleation of vortices.
In the system, the second and third terms in the right hand of Eq. (6) vanish since the velocity $\bm{v}$ is potential flow and a single-valued function of the position. 
Unless a low density region causing the divergence of the quantum pressure is formed at a point along the contour $C$, $\tilde{\mu}$ in Eq. (6) does not diverge on the $C$, which 
leads to $D\kappa /Dt = 0$. In this case, vortices cannot be nucleated because of Kelvin's theorem on circulation. 
However, $D\kappa /Dt$ does not generally vanish if a low density region causing the divergence of the quantum pressure is formed on $C$. 
This allows nucleation of the vortices, but the divergence is only the necessary condition for the nucleation of the vortices \cite{nucleation}. 

\begin{figure}[!t]
\begin{center}
\includegraphics[keepaspectratio, width=8cm,clip]{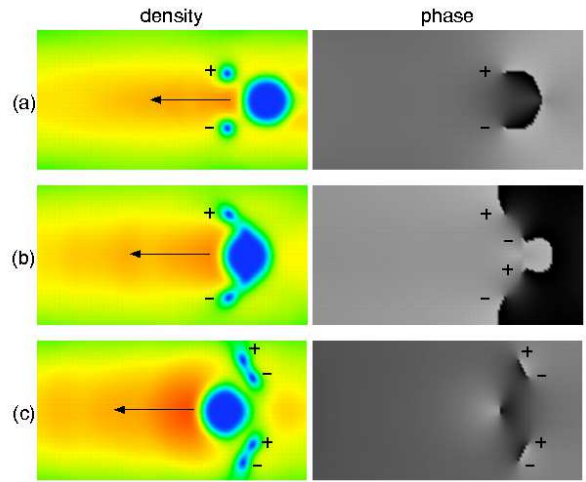}
\caption{(Color online) Reconnection of vortex pairs near the oscillating potential: The density (left) and phase (right) profiles at (a) $t$ = 31.4 ms, (b) $t$ = 38.3 ms , and (c) $t$ = 43.2 ms are shown, where the $x$ and $y$ dimensions of each image are 27.0 $\mu$m and 14.0$ \mu$m. The symbols $-$ and $+$ denote a vortex with clockwise or counterclockwise circulation, respectively. The black arrows indicate the direction of motion of the potential. The value of the phase varies from $-\pi$ (white) to $\pi$ (black). The density and phase profiles before the collision between the potential and the vortex pair are shown in (a). Thereafter, another ghost vortex pair nucleates in (b), exiting the potential through the collision, which causes reconnection of the vortices. As a result, two pairs appear in (c).  }
\end{center}
\end{figure}

This situation is directly described in our numerical simulation. The low density region in the condensate appears through two patterns. One is that the oscillating potential forms the low 
density region because of repulsion, which causes nucleation of vortices. The other is the inside of the solitons, which allows the transformation between vortices and solitons. 

\section{SYNERGY DYNAMICS OF VORTICES AND SOLITONS} 
An oscillating potential creates vortex pairs, causes reconnection of pairs characterized by the oscillation, and causes the new pairs to leave for the surface of the condensate.  
Consequently, the surface becomes filled with vortices having positive and negative circulation, which leads to nucleation of solitons and the migration of vortices. 
We call this sequence synergy dynamics of vortices and solitons, which often occurs in cases where the amplitude of the oscillation is larger than $L$. We demonstrate these dynamics through a simulation with $\epsilon = 10 \rm \mu m$ and  $ \omega = 60 \rm /s $. 
The overall dynamics have been briefly reported in \cite{s}, but this section describes each process in more detail. 

\begin{figure}[!t]
\begin{center}
\includegraphics[keepaspectratio, width=8cm,clip]{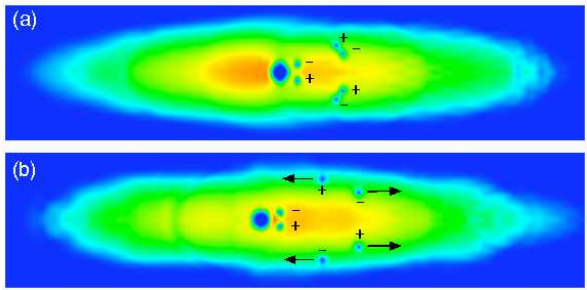}
\caption{(Color online) Separation of vortex pairs near the surface of the condensate: The density profiles at (a) $t$ = 58.9 ms and (b) $t$ = 88.4 ms are shown, where the $x$ and $y$ dimensions of each image are 145 $\mu$m and 34.0 $\mu$m, respectively. The symbols $-$ and $+$ denote a vortex with clockwise or counterclockwise circulation, respectively. The black arrows indicate the direction of motion of the vortices. 
The vortex pairs approach the surface of the condensate in (a) and separate and leave for the bow of the condensate  in (b).  }
\end{center}
\end{figure}

\begin{figure}[!b]
\begin{center}
\includegraphics[keepaspectratio, width=8cm,clip]{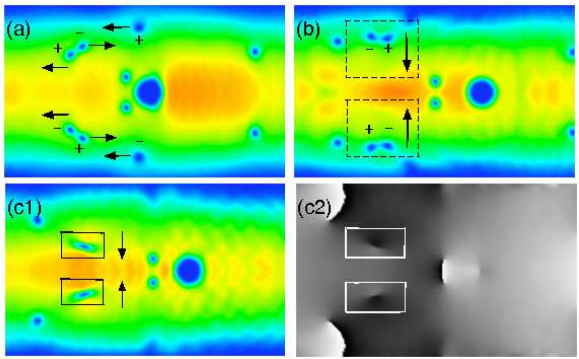}
\caption{(Color online) Nucleation of solitons: 
The density and phase profiles at (a) $t$ = 110 ms, (b) $t$ = 128 ms, and (c1)--(c2) $t$ = 138 ms are shown, where the $x$ and $y$ dimensions of each image are 45.2 $\mu$m and 27.6 $\mu$m, respectively. (c2) is the phase profile corresponding to the density profile (c1). The symbols $-$ and $+$ denote a vortex with clockwise or counterclockwise circulation, respectively. The black arrows indicate the direction of motion of the vortices and solitons. 
The value of the phase varies from $-\pi$ (white) to $\pi$ (black). Some vortices sit near the surface in (a) and reconnection of the vortices occurs in (b), where the square boxes with dotted lines enclose the new vortex pairs. While the new pairs move toward the center of the condensate, the pairs annihilate, which leads to nucleation of solitons in (c1) and (c2), shown by the square boxes with solid lines. }
\end{center}
\end{figure}

Ghost vortices, namely quantized vortices in a low density region, are important for the nucleation of the usual vortices  in the bulk density region since 
nucleation requires seeds of vortices. In rotating BECs, ghost vortices are nucleated outside the condensate, entering it through the excitation of the surface 
waves, leading to the creation of usual vortices \cite{Tsubota02,Kasamatsu03}. Thus, the periphery of the condensate provides seeds of topological defects. In our system, the oscillating potential provides seeds within itself. The potential starts to move, inducing a velocity field like back-flow, emitting phonons, and a ghost vortex pair is nucleated inside the potential as shown in Fig. 2(a). The ghost pair tends to move away from the potential in Fig. 2(b), and a usual vortex pair appears in the condensate in Fig. 2(c). Thus, the ghost vortices work as seeds of usual vortices. 

\begin{figure}[!b]
\begin{center}
\includegraphics[keepaspectratio, width=8cm,clip]{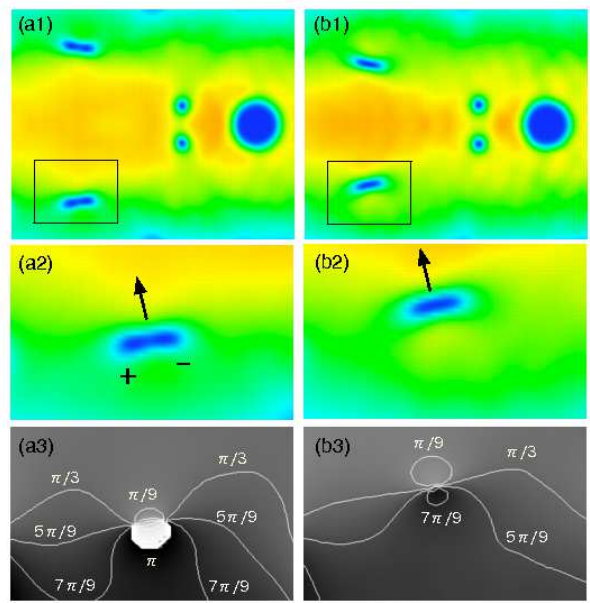}
\caption{(Color online) Structure of density and phase for nucleation of solitons: 
The density and phase profiles at (a1)--(a3) $t$ = 132 ms and (b1)--(b3) $t$ = 135 ms are shown. The $x$ and $y$ dimensions of (a1), (b1) are 27.0 $\rm \mu$m and 22.0 $\rm \mu$m, respectively, and for the rest are 13.0 $\rm \mu$m and 8.01 $\rm \mu$m, respectively. (a2) and (b2) show enlarged figures, depicted by square boxes with solid lines in (a1) and (b1). 
(a3) and (b3) are the phase profiles corresponding to the density profiles (a2) and (b2). 
(a1)--(a3) and (b1)--(b3) show the density and phase profiles before and after the annihilation of a vortex pair, respectively. 
The symbols $-$ and $+$ denote a vortex with clockwise or counterclockwise circulation, respectively. The black arrows indicate the direction of motion of the vortices and solitons. The value of the phase varies from $-\pi$ (white) to $\pi$ (black). (a1)--(a3) show the density and phase profiles before the annihilation of vortices, and the profiles after the annihilation are shown in (b1)--(b3).}
\end{center}
\end{figure}

Reconnection of vortex pairs occurs near the oscillating potential. The new vortex pair has an impulse in the same direction 
as that of the potential. Then, the potential changes the direction of the velocity. Thus, the potential will 
collide with the pair in Fig. 3(a). Then, a new ghost vortex pair is nucleated inside the potential whose impulse is opposite to that of the usual vortex 
pair, reconnecting with it as shown in Fig. 3(b). Thus, two new vortex pairs appear in the condensate in Fig. 3(c). Thereafter, the pairs move away from 
the potential, leaving for the surface of the condensate. This reconnection is characteristic of the oscillating potential because the 
potential repeatedly emits vortices of positive and negative circulation in opposite directions, which is not seen for potentials of uniform motion. 
This leads to nucleation of solitons, as shown in the following. 

\begin{figure}[!t]
\begin{center}
\includegraphics[keepaspectratio, width=8cm,clip]{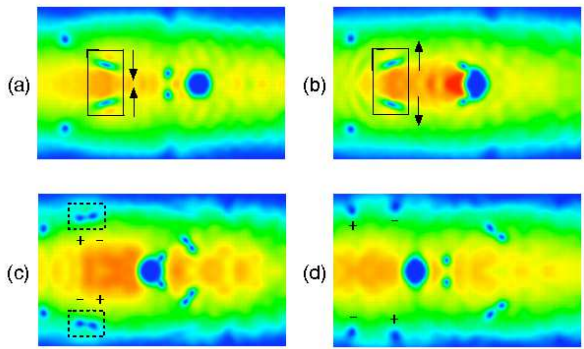}
\caption{(Color online) Collision and collapse of solitons: 
The density profiles at (a) $t$ = 139 ms, (b) $t$ = 147 ms , (c) $t$ = 157 ms, and (d) $t$ = 167 ms are shown, where the $x$ and $y$ dimensions of each image are 45.2 $\mu$m and 27.6 $\mu$m, respectively. The symbols $-$ and $+$ denote a vortex with clockwise or counterclockwise circulation, respectively. The black arrows indicate the direction of motion of the solitons. 
The square box with solid lines encloses the solitons and the box with dotted lines encloses the vortex pair. (a) and (b) show the state before and after the collision of the solitons, respectively. The decay of solitons to vortex pairs is shown in (c). (d) shows the density profile after the decay.}
\end{center}
\end{figure}

The vortex pairs separate as they approach the surface of the condensate in Fig. 4. This behavior is qualitatively understood by 
applying the idea of an image vortex, which is often used in hydrodynamics. The vortices induce a circular velocity field in a uniform 
system, but the field is distorted in a nonuniform system. This effect is strongly evident near the surface of the 
condensate where the density profile rapidly varies. As the vortices arrive at the surface, the normal component of the velocity field is suppressed. 
This situation is approximately equal to the relation between a vortex and a solid wall, so that the dynamics of vortices near the surface 
in Fig. 4 can be shown by the image vortex \cite{image}. Note that this idea only gives a qualitative understanding since the surface is not exactly a solid wall.  

The vortices near the surface of the condensate have two fates.
One is that a vortex pair transforms into solitons through the annihilation of the pairs. The other is that the vortices migrate in the condensate.
We show these dynamics in the following. 

\begin{figure}[!t]
\begin{center}
\includegraphics[keepaspectratio, width=8cm,clip]{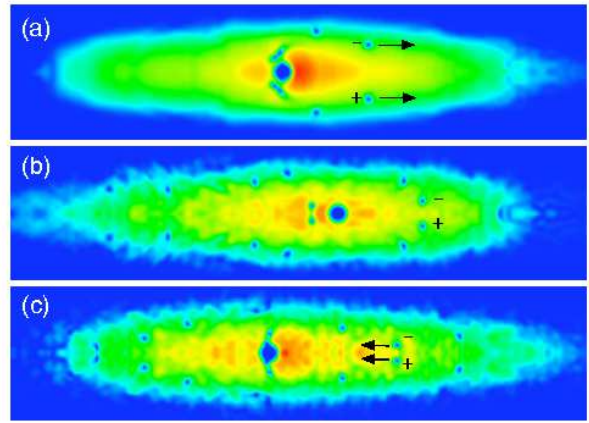}
\caption{(Color online) Migration of vortices: The density profiles at (a) $t$ = 98.2 ms, (b) $t$ = 236 ms , and (c) $t$ = 301 ms are shown, where the $x$ and $y$ dimensions of each image are 145 $\mu$m and 34.0 $\mu$m, respectively. The symbols $-$ and $+$ denote a vortex with clockwise or counterclockwise circulation, respectively. The black arrows indicate the direction of motion of the vortices. 
The vortices near the surface of the condensate in (a) move toward the bow, reconnecting there in (b) and returning to the center in (c). }
\end{center}
\end{figure}

{\it Transformation between vortices and solitons} {\it\Large-}
Many vortices have accumulated near the surface of the condensate in Fig. 5(a) since the oscillating potential continues to make vortices with positive and negative 
circulation. Hence, the vortices near the surface can reconnect with each other as shown in Fig. 5(b), where we enclose the new vortex pairs with 
square dotted lines. These pairs have impulse toward the center of the condensate. As a pair approaches the center, the 
size of the pair diminishes. Consequently, pair annihilation of vortices occurs, making the solitons shown in Figs. 5(c1) and (c2). 

A soliton of the GP equation with repulsive interaction has a locally decreased density profile 
and a rapidly varying phase profile. The low density parts in Figs. 5(c1) and (c2) have these properties and hence we can identify them as solitons.
This kind of nucleation of solitons is characteristic of an oscillating potential since it is caused by the potential emitting 
vortices in opposite directions. 

One may wonder why the annihilation of a vortex pair creates a soliton. 
Certainly, this would not occur if the system is uniform. However, our system is confined by a trapping potential, so that 
phenomena characteristic of a nonuniform system can occur, which is shown in Fig. 6. Figures 6 (a1) and (b1) show the density profiles immediately before and 
after the nucleation of solitons. 
We pay attention to the square boxes in Figs. 6(a1) and (b1), which are enlarged in Figs. 6(a2) and (b2). 
The phase profiles corresponding to Figs. 6(a2) and (b2) are Figs. 6(a3) and (b3), where the phase profiles are 
distorted because the vortices are in a narrow region between the high density region of the condensate and the surface.
Fig. 6(a3) shows that the phase of the vortex pair in the front and rear is flat while the phase in the right and left regions changes rapidly. 
In addition, the phase between the pair also changes more rapidly than that in the right and left regions. 
Then, the velocity is suppressed in the front and rear, and enhanced in other regions, where the velocity between the pairs is 
larger than that in the right and left regions. 
Therefore, the vortices attract each other through the Magnus force, and annihilation of the pair occurs, leading to the distorted phase shown in Fig. 6(b3) where 
the phase in the center of the low density region in Fig. 6(b2) rapidly changes by about $2 \pi /3$. 

\begin{figure}[!b]
\begin{center}
\includegraphics[keepaspectratio, width=8cm,clip]{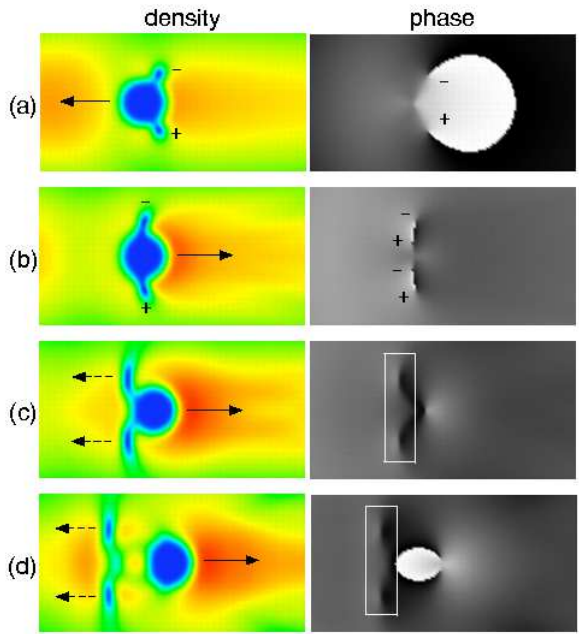}
\caption{(Color online) Nucleation of solitons near the oscillating potential: 
The density (left) and phase (right) profile at (a) $t$ = 23.6 ms, (b) $t$ = 27.5 ms , (c) $t$ = 30.3 ms, and (d) $t$ = 33.2 ms are shown, where the $x$ and $y$ dimensions of each image are 27.0 $\mu$m and 14.0 $\mu$m, respectively. The symbols $-$ and $+$ denote a vortex with clockwise or counterclockwise circulation, respectively. The black arrow with a solid line indicates the direction of motion of the potential, and 
the dotted arrow indicates the direction of motion of the solitons. 
The square boxes with solid lines enclose the solitons. The value of the phase varies from $-\pi$ (white) to $\pi$ (black). 
The amplitude and frequency of the oscillation are $3 \hspace{0.5mm} \mu \rm m$ and $190 \rm /s$, respectively.
(a) shows the density and phase profiles before the collision. Then, a new ghost vortex pair nucleates in (b), coming out of the potential. The annihilation of vortices leads to the nucleation of solitons in (c). These solitons move away from the potential in (d). }
\end{center}
\end{figure}

After the nucleation of solitons, the solitons collide and collapse. As shown in Fig. 7(a), the solitons move toward 
the center of the condensate, so that collision occurs as shown in Fig. 7(a), (b), which show the density profile before and after the collision. 
The solitons do not change their configurations during the collisions, which is a property of solitons. Thereafter they 
move towards the surface of the condensate, decaying into new vortex pairs as shown in Fig. 7(c). 
At the surface, the pairs separate and the resulting vortices move toward the bow of the condensate by the same mechanism in Fig .4.
Note that these dynamics do not always occur, because the oscillating potential occasionally causes the solitons to crash, hindering the above dynamics. 

\begin{figure}[!t]
\begin{center}
\includegraphics[keepaspectratio, width=8cm,clip]{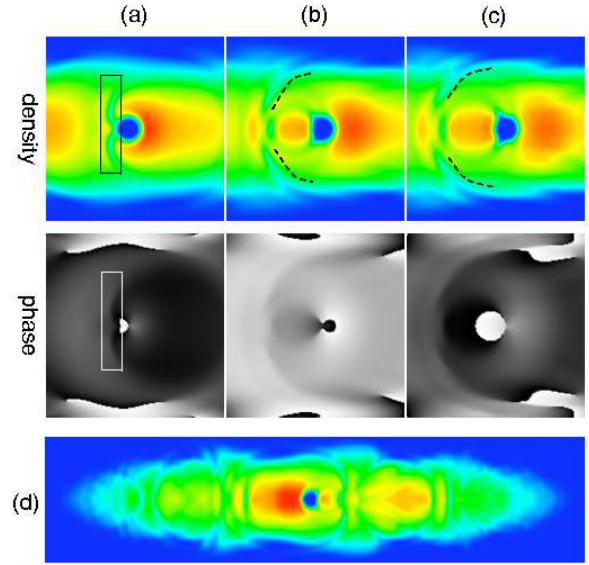}
\caption{(Color online) Transfer of solitons: 
The density and phase profiles at (a) $t$ = 20.8 ms, (b) $t$ = 25.5 ms , and (c) $t$ = 27.1 ms are shown, where the $x$ and $y$ dimensions of each image are 32.8 $\mu$m and 34.0 $\mu$m, respectively. The overall density profile at t = 59.0 ms is shown in (d), where the $x$ and $y$ dimensions of each image are 145 $\mu$m and 34.0 $\mu$m, respectively. The symbols $-$ and $+$ denote a vortex with clockwise or counterclockwise circulation, respectively. The value of the phase varies from $-\pi$ (white) to $\pi$ (black). The amplitude and frequency of the oscillation are $2 \hspace{0.5mm} \mu \rm m$ and $280 \rm /s$, respectively. 
The solitons, indicated by the square box with solid lines, nucleate by the potential in (a), moving toward the surface of the condensate in (b)--(c), where the dotted lines show the solitons. 
The solitons cut on  the condensate in (d). }
\end{center}
\end{figure}

{\it Migration of vortices} {\it\Large-}
The vortices that do not change into solitons migrate in the condensate in Fig. 8 where two vortices among them are depicted by $+$ and $-$ symbols. First, the vortices near the surface of the condensate move towards the bow as shown in Fig. 8(a). Second, the vortices that have reached the bow reconnect with each other in Fig. 8(b). Third, the new vortex pair returns to the center of the condensate in Fig. 8(c). Figure 8 also shows that the surface is distorted by the phonons emitted by the oscillating potential. 

Thus, the oscillating potential leads to the synergy dynamics and migration of vortices. Migration is also observed in a uniformly moving potential \cite{Neely10}, 
but other dynamics are characteristic of an oscillating potential, which cannot occur for a uniformly moving potential. 

\begin{figure}[!t]
\begin{center}
\includegraphics[keepaspectratio, width=8cm,clip]{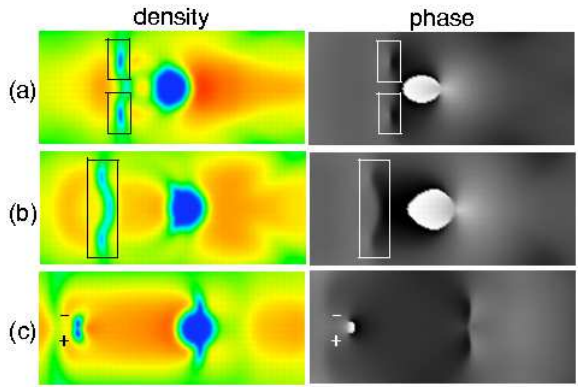}
\caption{(Color online) Soliton decay in the center of the condensate: The density (left) and phase (right) profile at (a) $t$ = 33.2 ms, (b) $t$ = 36.3 ms, and (c) $t$ = 42.4 ms are shown, where 
the $x$ and $y$ dimensions of each image are 32.8 $\mu$m and 14.0 $\mu$m, respectively. The symbols $-$ and $+$ denote a vortex with clockwise or counterclockwise circulation, respectively. 
The value of the phase varies from $-\pi$ (white) to $\pi$ (black).
The amplitude and frequency of the oscillation are $3 \hspace{0.5mm} \mu \rm m$ and $190 \rm /s$, respectively. (a)--(b) show the solitons moving away from the potential, which are depicted by the square boxes with solid lines. The soliton decays to a vortex pair in (c). }
\end{center}
\end{figure}

\section{NUCLEATION OF SOLITONS NEAR THE OSCILLATING POTENTIAL}
An oscillating potential with an amplitude smaller than $L$ leads to different dynamics than the case of a large amplitude. The most distinctive 
dynamics occur in the collision between vortex pairs and the oscillating potential. In the following, we show the dynamics related to the 
collision between the pairs and the potential, and the propagation and decay of solitons. 

Nucleation of solitons near the oscillating potential occurs in the case of a small amplitude, shown in Fig. 9. The mechanism 
of vortex nucleation in Fig. 9(a) is the same as the case for a large amplitude. The difference occurs in the collision between vortex pairs and the oscillating 
potential in Fig. 9(b). The distance between the vortices and the potential is closer than that in the case for a large amplitude, which means that 
vortices in the condensate are close to the ghost vortices inside the potential. Therefore the annihilation of vortices occurs easily, leading to the nucleation of 
solitons in Fig. 9(c). Thereafter, these solitons move away from the potential in Fig. 9(d).

\begin{figure}[!b]
\begin{center}
\includegraphics[keepaspectratio, width=9cm,clip]{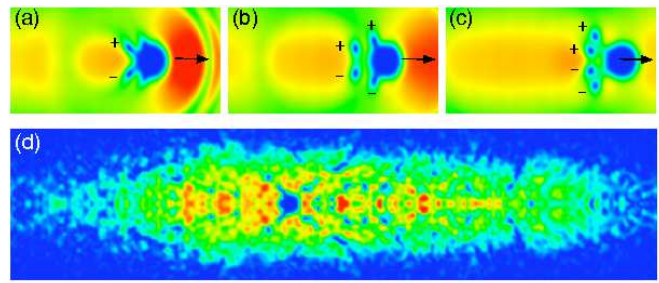}
\caption{(Color online) Case for high velocity: 
The density profiles at (a) $t$ = 5.89 ms, (b) $t$ = 8.84 ms, and (c) $t$ = 13.7 ms are shown, where the $x$ and $y$ dimensions of each image are 28.0 $\mu$m and 14.0 $\mu$m, respectively. 
The overall density profile at $t$ = 265 ms is shown in (d), where the $x$ and $y$ dimensions of image are 145 $\mu$m and 34.0 $\mu$m, respectively. The symbols $-$ and $+$ denote a vortex with clockwise or counterclockwise circulation, respectively. The black arrows indicate the direction of motion of the potential. The amplitude and frequency of the oscillation 
are $10 \hspace{0.5mm} \mu \rm m$ and $90 \rm /s$, respectively. A vortex pair is nucleated in (a) and the potential nucleates another pair in (b). 
As a result, multiple vortex pairs in (c) appear in the condensate. The potential continues nucleating multiple pairs, which leads to the distorted condensate in (d). }
\end{center}
\end{figure}

This phenomenon is attributed to the relation between the nucleation time and the quarter period of the oscillation.
Generally, it takes a time $t_{v}$ for a moving potential to nucleate a vortex pair. If the time $t_{v}$ is larger than the period 
$t_{p} = \pi /2\omega$, the nucleation cannot occur since the velocity of the potential decreases after the period $t_p$ before the time $t_v$ passes. 
In the case of a large amplitude, $t_{p}$ of the oscillation is much longer than $t_{v}$ because the 
frequency $\omega$ of the oscillation is small. 
Thus, the potential can move in one direction for a long time $t_{p} > t_{v}$, nucleating the vortices. After the 
nucleation, the potential still moves in the same direction, so that the vortices part from it as shown in Fig. 2(c). On the other hand, a 
potential with a small amplitude needs a large frequency to cause nucleation, which implies that the potential can move for a short time $t_{p} \geq t_{v}$. 
In this case, the vortices are nucleated just before the potential changes the direction of the motion, as shown in Figs. 9(a) and (b). Thus, the 
situation in which the distance between the vortices and the potential is close causes annihilation of vortices near it, and the solitons
are nucleated in Fig. 9(c). 

\begin{figure*}[!t]
\begin{center}
\includegraphics[keepaspectratio, width=8cm,clip]{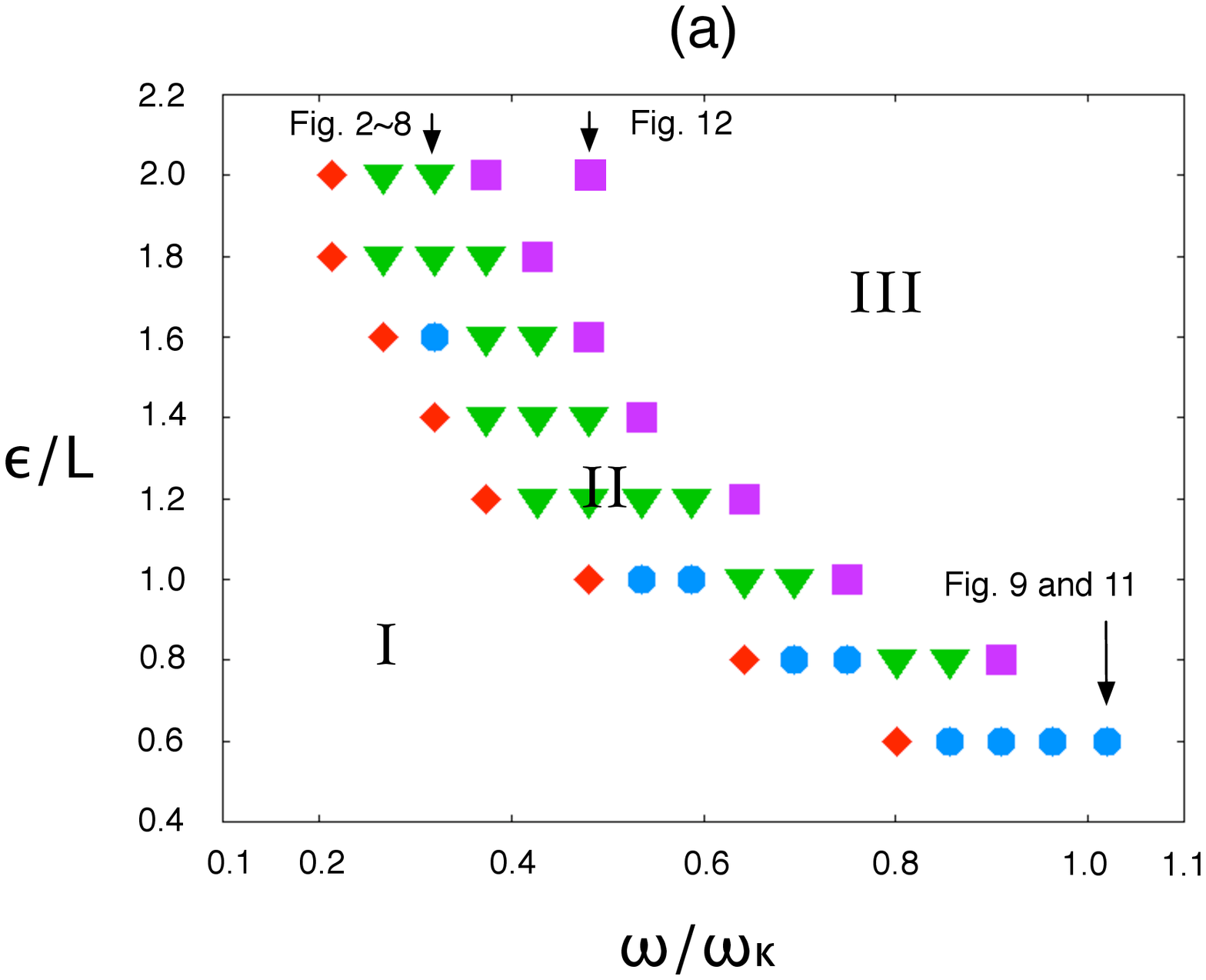}
\includegraphics[keepaspectratio, width=8cm,clip]{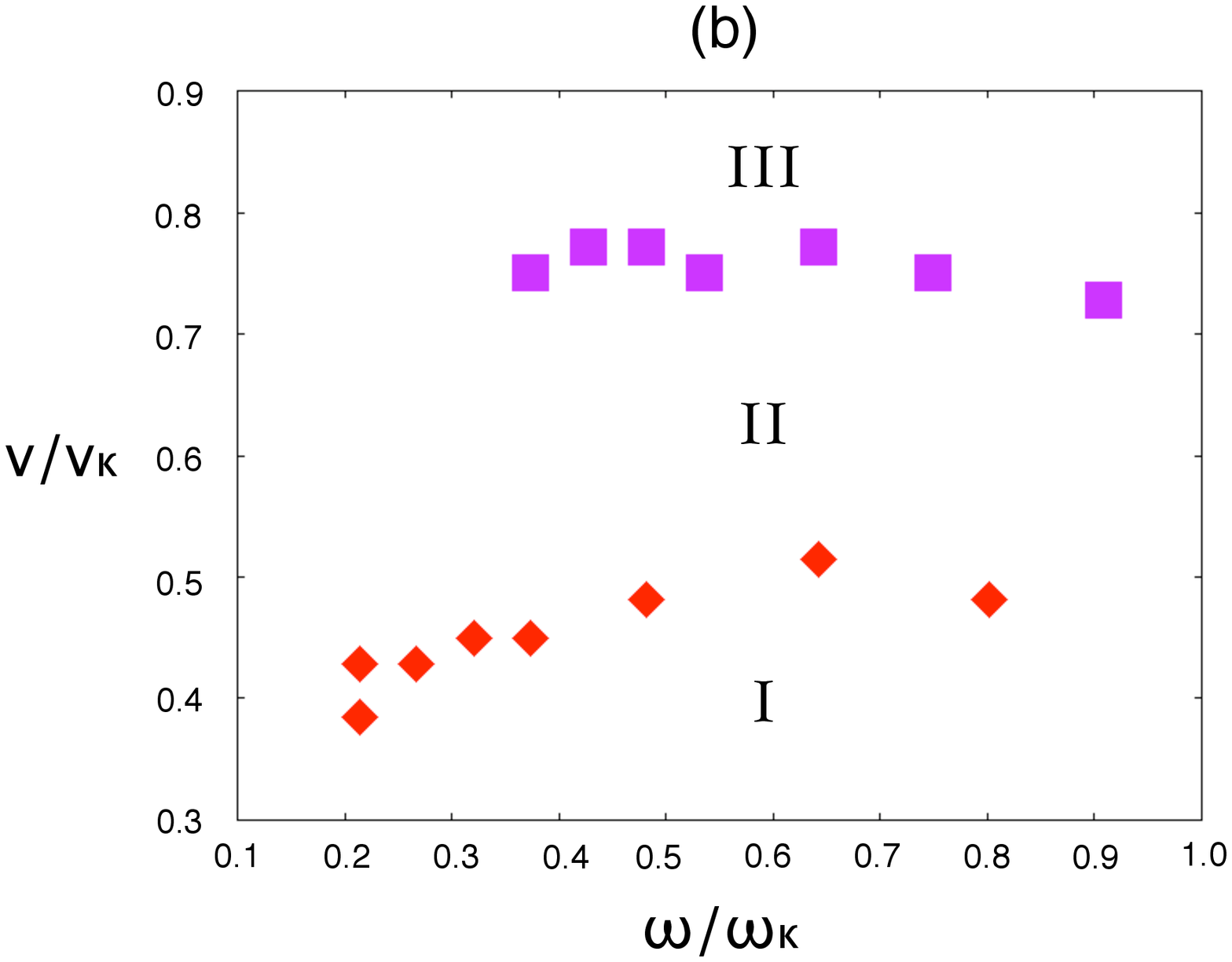}
\caption{(Color online) Phase diagram of the nucleation of vortices and solitons by an oscillating potential: (a) shows the $ \epsilon $--$ \omega $ phase diagram of the nucleation. 
In (a) there are four kinds of points: no nucleation of vortices and solitons ($\blacklozenge$), reconnection of vortices near the potential ($\blacktriangledown$) as 
described in Fig. 3, nucleation of solitons near the potential ($\bullet$) as described in Fig. 9 and nucleation of multiple vortex pairs ($\blacksquare$) as described in Fig. 12. 
(b) shows the $ v $--$ \omega $ phase diagram, which is related to the $ \epsilon $--$ \omega $ phase diagram through $v=\epsilon \hspace{0.5mm} \omega$.
The critical velocity for the nucleation of a vortex pair and multiple vortex pair is illustrated by two kinds of points in (b): no nucleation of vortices 
and solitons ($\blacklozenge$) and nucleation of multiple vortex pairs ($\blacksquare$).}
\end{center}
\end{figure*}

There are two patterns for the dynamics of solitons after nucleation. One is that the solitons move toward the surface without breaking, and the other is that they decay to a vortex pair in the center of the condensate.

We show the former pattern in Fig. 10, where the amplitude and frequency of the oscillation are $2 \hspace{0.5mm} \mu \rm m$ and $280 \rm /s$.
The solitons are nucleated near the potential in Fig. 10(a) through the mechanism in Fig. 9. In this case, the solitons move away from the oscillating potential and 
toward the surface of the condensate, making cuts in it as shown in Fig. 10(b) and (c), where the dotted lines show the solitons.   
The oscillating potential keeps nucleating solitons moving toward the surface, so that several cuts are made in Fig. 10(d) 
in contrast to the case for large amplitude in Fig. 8(c). 

We show the latter pattern, in which the solitons decay to a vortex pair in the center of the condensate, 
in Fig. 11, where the amplitude and frequency of the oscillation are $3 \hspace{0.5mm} \mu \rm m$ and $190 \rm /s$, respectively. 
As shown in Fig. 11(a) and (b), the solitons are nucleated near the oscillating potential, shown enclosed by the square boxes. 
Then, a vortex pair leaving for the bow is nucleated in the center of the condensate as shown in Fig. 11(c). Thereafter, when the pair arrives at the bow, it separates and moves 
along the surface. 

\section{NUCLEATION AND DYNAMICS OF MULTIPLE VORTEX PAIRS}

We show the dynamics for high velocity, which means that the velocity is much higher than the critical velocity for nucleation of vortices discussed later. 
The higher the velocity of the potential, the shorter the nucleation time. Thus, it is possible for the potential to make multiple vortex pairs in Figs. 12(a)--(c), where the amplitude 
and frequency of the oscillation are $10 \hspace{0.5mm} \mu \rm m$ and $90 \rm /s$, respectively. 
After the nucleation of multiple vortex pairs, the potential collides with them. In this case, both reconnection of vortices and nucleation of solitons 
can occur near the potential. As a result, the condensate is filled with vortices and solitons in Fig. 12(d).

The boundary of the condensate in Fig. 12(d) shows the sign of the granulation of the condensate \cite{granulation}. If a potential with high velocity would oscillate for a long time, the whole 
condensate could become granular. 

\section{PHASE DIAGRAM OF NUCLEATION OF VORTICES AND SOLITONS BY AN OSCILLATING POTENTIAL}

We now consider the nucleation of vortices and solitons by the oscillating potential and systematically investigate the dependence of the nucleation on the parameters. 
To make a phase diagram for the nucleation,  we consider what dimensionless parameters are characteristic of the system. 
Our system has two features: it is a quantum fluid and is subject to external oscillation. 
The first feature is represented by quantum circulation $ \kappa = h/m $, from which we define the characteristic velocity $v_{\kappa} = h/mL$ and 
the characteristic frequency $\omega _{\kappa} = h/mL^2$. 
The second feature is represented by the Strouhal number \cite{Landau} peculiar to the oscillating field, which is expressed by $V_{s} \tau _{s} / L_{s}$, where $V_{s}$, $\tau _{s}$, and $L_{s}$ are the velocity, time, and length characteristic of the system. In our case, the Strouhal number is $\epsilon / L$ for $V_{s} = \epsilon \hspace{0.5mm} \omega$, $\tau _{s} = 1/ \omega$ and $L_{s} = L$. 
Then, we create the phase diagram of the nucleation with three dimensionless parameters $\epsilon / L$, $\omega / \omega _{\kappa}$ and $ v / v_{\kappa}$. 
Since these are related to each other through $v = \epsilon \hspace{0.5mm} \omega$, there remain two independent parameters. Figure 13 shows the phase diagram obtained 
by the numerical calculation. The $ \epsilon $--$ \omega $ phase diagram is shown in Fig. 13(a), where there are four kinds of points that denote no nucleation 
of vortices and solitons ($\blacklozenge$), nucleation of vortices and reconnection of vortices near the potential ($\blacktriangledown$) 
as described in Fig. 3, nucleation of solitons after the nucleation of vortices near the potential ($\bullet$) as described 
in Fig. 9, and nucleation of multiple vortex pairs ($\blacksquare$) as described in Fig. 12. 
Note that nucleation of vortices means that the oscillating potential nucleates usual vortices, discussed in the next chapter in detail. 
The velocity $v$ of the potential is expressed 
by $ \epsilon \hspace{0.5mm} \omega $, so that we can transform the $ \epsilon $--$ \omega $ phase diagram into the $ v $--$ \omega $ phase diagram in Fig. 13(b) to show the dependence of the critical velocity for the nucleation of a vortex pair and a multiple vortex pair on $\omega$.

The phase diagram shows that nucleation induced by the oscillating potential can be roughly classified into three regions. In region I, vortices and solitons are never nucleated by the 
potential, which only emits phonons. 
In region II, the potential nucleates a vortex pair, which is described in Secs. IV and V. Besides, region II can be classified 
into two sub-regions. One is a sub-region where reconnection of vortex pairs occurs near the potential, as shown in Fig. 3. 
In the other sub-region solitons are nucleated near the potential, as shown in Fig. 9. 
Figure 13(a) shows that the region for reconnection of vortices near the potential  ($\blacktriangledown$) is wider than that for nucleation of solitons near the potential ($\bullet$) if the 
amplitude $\epsilon$ is larger than $L$. 
The region corresponding to $\epsilon /L = 1.2 \sim 2.0 $ is mostly filled with $\blacktriangledown$ symbols, but 
we observe nucleation of solitons ($\bullet$) for $\epsilon /L = 1.6$ because of the following. 
The distance between the vortices and the potential is essential for the dynamics ($\bullet$) as explained in Sec. V. 
When the amplitude and frequency of the potential is slightly larger than the critical values for nucleation of vortices, the distance is small. 
Therefore, nucleation of solitons should occur even for large amplitude if the parameters are close to the critical value.  
However, the region of this nucleation is narrow if the amplitude is larger than $L$. 
Thus, the $\bullet$ point with $\epsilon / L = 1.6$ appears.
The region of the $\bullet$ symbol should survive even for a  large amplitude ($\epsilon > L$) if we finely sweep the frequency and amplitude.
In region III, multiple vortex pairs are nucleated and the complex dynamics of vortices and solitons as described in Sec. VI is induced by the potential.

\begin{figure*}[!t]
\begin{center}
\includegraphics[keepaspectratio, width=8cm,clip]{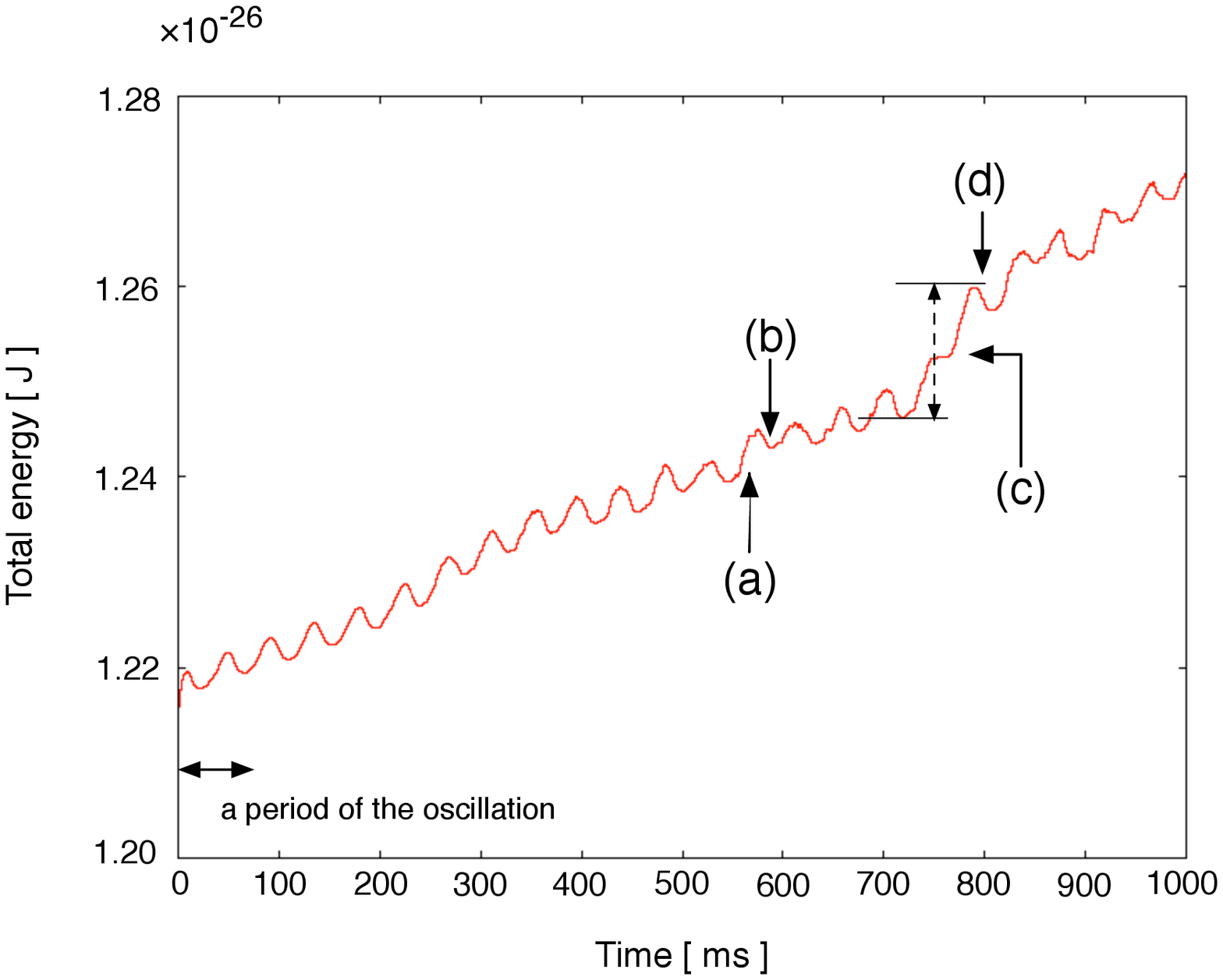}
\includegraphics[keepaspectratio, width=8cm,clip]{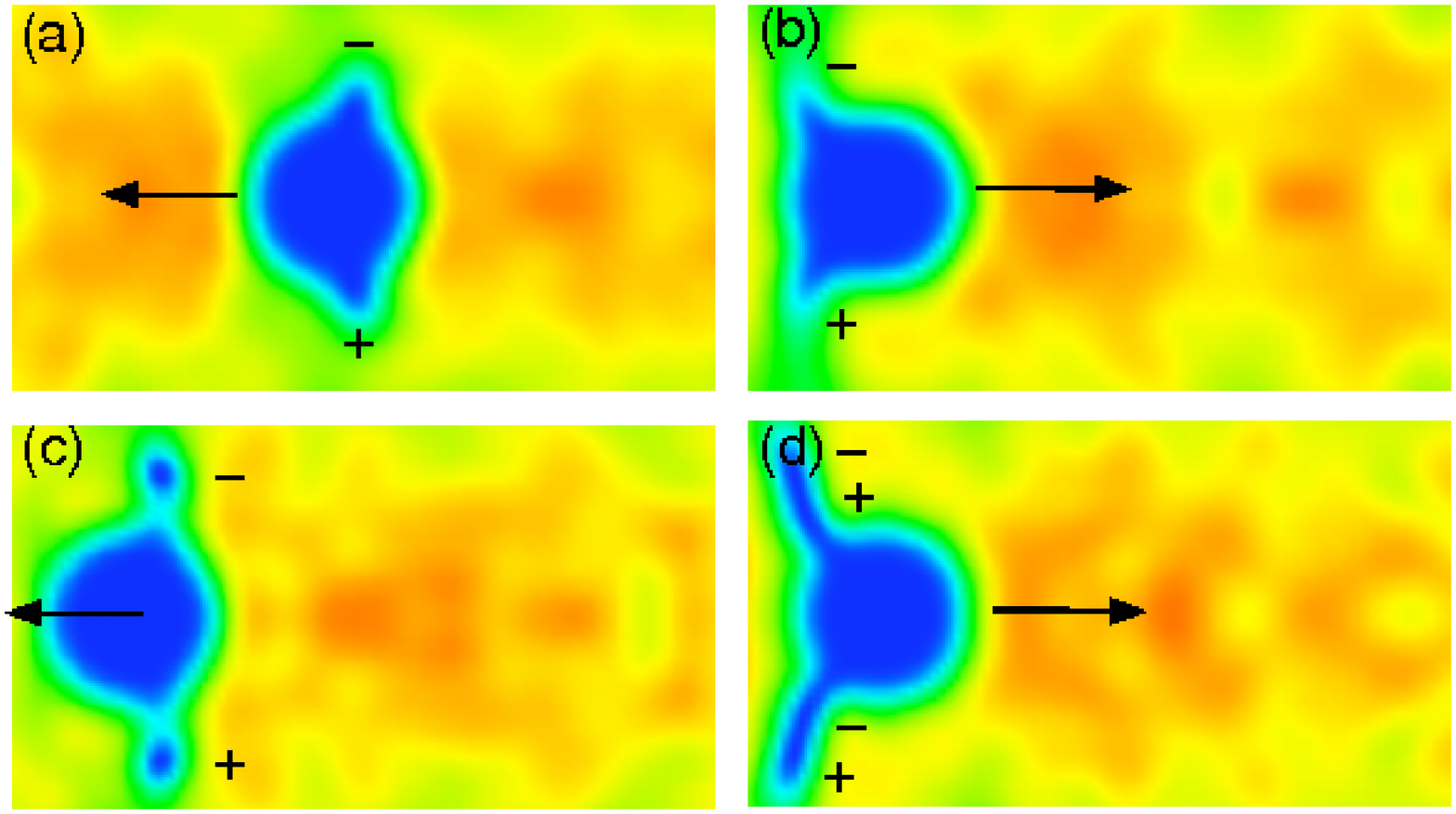}
\caption{(Color online) The increase of the total energy in the case with $\epsilon=6$ $\mu$m and $\omega = 73.4$ $/s$ is shown.
(a)--(d) show the density profiles at $t$ = 562, 580, 745, 758 ms. The symbols $-$ and $+$ denote a vortex with clockwise or counterclockwise circulation, respectively. The black arrows indicate the direction of motion of the potential. The energy graph shows that there is a large increase for 750--850 ms, corresponding to the energy of vortices. 
(a)--(b) show density profiles in which the vortices appear only within the boundary layer, not those distant from the potential. 
Vortices appear in the condensate in (c)--(d), which is regarded as the nucleation of vortices. } 
\end{center}
\end{figure*}

It is not the case that the dynamics induced by an oscillating potential approximate those of a uniformly moving potential if the frequency $\omega$ approaches zero.
This is because, even though the potential oscillates with low frequency, it nucleates vortex pairs with different 
charges, and the reconnection of vortices or the nucleation of solitons near the potential not caused by the uniformly moving potential can occur. 
Therefore, the dynamics under an oscillating potential is qualitatively different from that with a uniformly moving potential. 

\section{CRITICAL VELOCITY FOR NUCLEATION OF VORTICES}
The mechanism of nucleation of quantized vortices is one of the most important issues for quantum fluids. 
This is because the vortices, which are stable topological defects, bring remarkable changes to the system once they are nucleated. 
For example, the decay of the permanent flow of superfluid helium is considered to be strongly related to the motion of vortices, which leads to phase slippage. 

Our numerical calculations suggest that the mechanism of nucleation of vortices in a system with an oscillating potential is different from that 
in a system where the condensate flows uniformly without any obstacles. 
Nucleation in a uniform system requires excitation of phonons since a low density region causing divergence of the quantum pressure can 
be formed only by the growth of the amplitude of phonons, which leads to the breakdown of Kelvin's theorem on circulation, as explained in Sec. III. 

However, in a system with an oscillating potential, a different mechanism of nucleation applies. 
In this system, the potential creates a low density region inside itself, causing divergence of the quantum pressure. 
This leads to nucleation without any growth of phonons, as shown in Fig. 2 where a ghost vortex 
pair is nucleated inside the potential and moves away from it to become a usual vortex pair. 
According to the numerical calculations, an oscillating potential continues nucleating phonons in the condensate, though the phonons do not grow to create a low density region 
causing the divergence when the amplitude and frequency of the oscillation is near the critical value for nucleation of vortices. 
If the parameters of the oscillation are much larger than the critical values, the phonons nucleated by the potential may form a low density region. 
Therefore, our numerical calculations show that the critical velocity for nucleation of vortices by a potential is smaller than that by the growth of phonons in our system. 

We now consider the nucleation of a single vortex pair in Fig. 2, rather than multiple vortex pairs in Fig. 12, and define the nucleation induced by the oscillating potential. 
Our definition of nucleation is based on the increase of the total energy of the system and the density profile as shown in the following. 
The total energy increases by the energy of the vortices when the potential nucleates vortices, which is roughly estimated to be the order of $10^{-28}$ $\rm J$ in our simulation by 
using the energy of a vortex in a uniform system \cite{p} expressed by 
\begin{equation}
\epsilon _{v} = \pi n_{0} \frac{\hbar ^2}{m} {\rm{log}}(1.464 \frac{D}{\xi}), 
\end{equation}
where $D$ is the size of the system, which is the Thomas--Fermi radius $R_y$ in our case. 
Figure 14 shows the time evolution of the total energy $E_t$ expressed by 
\begin{equation}
E_{t} = \int \psi ^* (-\frac{\hbar ^2}{2m} \bm{\nabla} ^2) \psi d\bm{r} + \int V |\psi|^2 d\bm{r} + \frac{g}{2} \int |\psi|^4 d\bm{r}. 
\end{equation}
The total energy monotonically increases with oscillation because phonons are excited by the oscillating potential. 
For example, Fig. 14(a) and (b) show the density profile with a monotonic increase of the energy. 
In Fig. 14 (a), the vortices do not sufficiently escape from the boundary layer of the width of the order $\xi$ of the potential, and 
the increase of the total energy for the oscillation is smaller than $10^{-28}$ $\rm J$. 
After Fig. 14 (a), when the potential changes its direction of motion, ghost vortices, with a direction of impulse opposite to that of the usual vortices 
are nucleated inside the potential and the usual vortices disappear through annihilation with the ghost vortex pairs. 
Figure 14 (b) shows the density profile just before the annihilation. The vortices disappear within the boundary layer of the potential and no solitons are nucleated. 
Then, neither reconnection in Fig. 3 nor nucleation of solitons in Fig. 9 occur. 
Thus, vortices appear and disappear within the boundary layer of the potential. 
We judge that the vortices are not yet nucleated by the potential in this case. 
On the other hand, the behavior of the total energy in Fig. 14 changes markedly during $t=750$--$850$ $\rm ms$ when the vortices are nucleated by the oscillating potential. 
Figure 14 (c) shows the density profile for the nucleation of vortices, which shows that vortices sufficiently escape from the boundary layer of the potential. 
An increase of the order of $10^{-28}$ $\rm J$ occurs twice within a period of the oscillation $2 \pi / \omega$ because of the following. 
The potential moves in the $x$ direction and nucleates a vortex pair during the former half period $\pi / \omega$. Then, it starts to move back in the $-x$ direction and 
nucleates another pair which has an impulse opposite to that of the previous pair during the latter half period $\pi / \omega$. Thus, four vortices are nucleated in Fig. 14(d), which is peculiar to an oscillating potential. 
According to the above, we judge that vortices are nucleated if an increase of total energy of the order of $10^{-28}$ $\rm J$ occurs twice, as shown in Fig. 14. 
Using this criterion for nucleation, we find that the critical velocity of the vortices $v_{c}/v_{\kappa}$ is $0.385 \sim 0.514$ as shown in Fig. 13(b) ($\blacklozenge$) by changing the 
amplitude and the frequency of the oscillation by 1 $\rm \mu$m and 10 $\rm /s$, respectively. 

There are two candidates for the critical velocity for nucleation of vortices in our system. 
One is the sound velocity $c_{s}$ and the other is $v_{\kappa} = \kappa /L$. 
These velocities are based on the Landau instability, which occurs at $ v = {\rm{min}}[E_{e} / p_{e}]$, where $E_{e}$ and $p_{e}$ are the energy and momentum 
of an excitation, respectively. If the energy and momentum of a Bogoliubov excitation (a vortex pair) is chosen to be $E_{e}$ and $p_{e}$, respectively, we obtain $c_{s}$ ($v_{\kappa}$). 
In the following, we consider whether these velocities are the critical velocity in our system. 

The sound velocity $c_s$ near the potential is often compared with the critical velocity $v_c$ for nucleation of vortices in atomic BECs \cite{Jackson00,Neely10,Crescimanno2000,Huepe00}. 
However, the velocity $c_s$ cannot be the critical velocity $v_c$ in our system where a local potential moves in the condensate.  
The first reason is that the standard representation $c_{s}=\sqrt{g n_{0} /m}$ is rewritten as $c_s = \kappa /2 \sqrt{2} \pi \xi$ including 
the coherence length $\xi = \hbar /\sqrt{2m n_{0} g}$ and the quantum circulation $\kappa$, but the size $L$ of the potential is not included. 
Sasaki $et$ $al$.  \cite{Sasaki10} have reported that the critical velocity $v_c$ is dependent on $L$, which is obtained by the numerical simulation of the GP equation 
with a uniformly moving potential. 
The second reason is that the velocity $c_s$ is just the critical velocity for Landau instability related to phonon emission when the condensate flows in a uniform system with dissipation and noise. Our simulation shows that the vortices can be nucleated without the growth of phonons, which means that the velocity $c_s$ is not directly related to 
the nucleation of vortices, being just the critical velocity for which Bogoliubov phonons are excited in a uniform system. 
When the critical velocity $v_c / v_{\kappa}$ in Fig. 13(b) is renormalized by the sound velocity $c_{s}$, $v_{c}/c_{s}$ is about $0.292\sim 0.391$. 
A similar result has been reported by Jackson $et$ $al$.  \cite{Jackson00}. The disagreement between the critical velocity and the sound velocity is attributable to this discussion.

The velocity $v_{\kappa}$ cannot be the critical velocity $v_{c}$ either since the healing length $\xi$ is not included in the velocity $v_{\kappa}$. 
The coherence length can influence the critical velocity for nucleation of vortices in atomic BECs, which is different 
from the case of superfluid helium.  In helium, the critical velocity is considered to be almost dominated by the quantum 
circulation $\kappa$, from which the velocity is estimated to be $v_{\kappa}$ \cite{Pines}. 
This idea is based on the situation where the size of the oscillating object is much larger than the coherence length in helium; the 
size of the objects is usually larger than the order of $\mu$m and the coherence length is of the order of $\AA$. 
The small coherence length is caused by the strong interaction between helium atoms. 
However, in dilute atomic BECs, the coherence length is slightly smaller than the size of the oscillating potential because the system is dilute and the 
interaction between particles is weak. The typical coherence length and size of the potential are of the order of $0.1$ $\mu$m and $1$ $\mu$m, respectively. 
Thus, the critical velocity in atomic BECs does not conform with the rough estimate $v_{\kappa}$ in which the coherence length is not taken into account \cite{Zwerger00}.  
Our numerical calculations show that the critical velocity $v_{c}/v_{\kappa}$ is $0.385$--$0.514$ as shown in Fig. 13 ($\blacklozenge$), which is considerably smaller than the characteristic velocity $v_{\kappa}=934$ $\mu$m/s. 

From the above discussions, showing the importance of $\kappa$, $L$, and $\xi$ for the critical 
velocity, we propose an expression for the critical velocity including these variables: 
\begin{equation}
v_{c} = \frac{\kappa}{L + \alpha \xi},
\end{equation}
where $\xi$ is the coherence length near the center of the condensate and $\alpha$ is a parameter. In our case, $\alpha$ is about $11$--$19$ because the critical velocity 
is $360$--$480$ $\mu$m/s. 
This means that the potential is covered with a low density region of the width of the order of $\xi$, and the effective size of the potential is larger than $L$. 
This expression approaches $v_{\kappa}$ and $c_s$ for the two limiting cases. If $L$ is much larger than $\xi$, $v_{c}$ becomes $v_{\kappa}$. 
On the other hand, $v_{c}$ is close to the order of $c_{s}$ if $\xi$ is much larger than $L$, which means that the system becomes uniform. 
Note that this discussion should be correct for $V_{0} >> \mu$ because of the following. 
If the strength $V_0$ of the oscillating potential satisfies $V_{0} >> \mu$, an increase of $V_0$ only enlarges $L$ without changing the density 
profile inside the potential, where the density is very low for $ V_{0} >> \mu $. Therefore, Eq. (9) contains information on $V_0$ through $L$. 
However, if $V_0$ becomes comparable to $\mu$, the condensate penetrates into the potential. 
In this case, Eq. (9) may be insufficient since the density profile inside the potential changes dependently on $V_0$. 

In our case, the oscillating potential is treated, but Eq. (9) would also be valid in a system with a uniformly moving potential. 
Neely $et$ $al$.  \cite{Neely10} observed the critical velocity in a trapped system with a uniformly moving potential and found $v_{c} = 0.1 \hspace{0.5mm} c_{s}$. 
In their experiment, the healing length and size of the potential were $\xi = 0.3 \hspace{0.5mm} \rm \mu m$ and $L \sim 20 \hspace{0.5mm} \rm \mu m$, respectively. 
As Eq. (9) is applied to the result, the parameter $\alpha$ is about $21$. 
Moreover, we obtain $\alpha \sim 20$ when Eq. (9) applies to the critical velocity obtained by Sasaki $et$ $al.$  \cite{Sasaki10}, which 
is the numerical result of the GP equation with a uniformly moving potential.
Thus, $\alpha$ in Eq. (9) takes a similar value, $\alpha = 10$--$20$, being independent of whether the system has a uniformly moving potential or an oscillating potential. 
We consider that the value of the parameter $\alpha$ has the same order in these systems 
if the shape of the potential is the same and the strength of the potential is much larger than the chemical potential. 

\begin{figure}[!t]
\begin{center}
\includegraphics[keepaspectratio, width=8cm,clip]{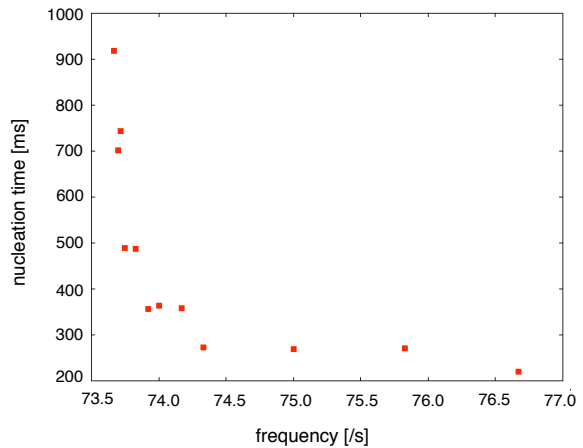} 
\caption{(Color online) Dynamical critical phenomena for nucleation of vortices: The relation between the nucleation time $t_{v}$ and frequency $\omega$ is shown. The amplitude of the 
oscillation $\epsilon$ is set to be $6 \hspace{0.5mm }\mu m$. The time $t_v$ diverges as the frequency decreases to approach $\omega _c = 73.5$ /s. } 
\end{center}
\end{figure}

\section{DIVERGENCE OF NUCLEATION TIME OF VORTICES}
We now address the issue of vortex nucleation with the relation of critical phenomena.
Customary critical phenomena are classified into two categories, static and dynamic.
Static phenomena concerning equilibrium properties mean that a system shows a characteristic power law for order parameters, susceptibility, specific heat, {\it etc.} near the critical region.
Dynamic phenomena concerns time dependent phenomena such as relaxation time, diffusion constant, {\it etc.}
Such dynamic critical phenomena have been reported in quantum fluid dynamics.
Thermal counterflow of superfluid $^4$He is known to experimentally show a transition from a laminar state to turbulence states consisting of a tangle of quantized vortices when we increase the relative velocity between the normal fluid and superfluid \cite{Tough}.
When the turbulence is generated in a tube of circular cross section, a vortex-line density is observed to develop from a low-density state (T-I) to a higher-density state (T-II) that can be associated with the homogeneous state.
Griswold {\it et al.} observed that a large fluctuation of the vortex-line density sharply increased the time constants of the system near the transition between T-I and T-II \cite{Griswold87}.  
The transition to QT in thermal counterflow can be understood in the light of a recent numerical simulation of the vortex filament model based on the full Biot--Savart law \cite{Adachi10}. The relaxation time to a turbulent state increases sharply near the critical region between the laminar and the turbulent states because of the large fluctuation of the vortex-line density \cite{Adachi}. 

The nucleation time $t_{v}$ of vortices by an oscillating potential diverges near the critical amplitude and frequency of the oscillation. 
We expect that there is a critical region where a power law between them  is realized because Huepe {\it et al}. \cite{Huepe00} considered a power law with saddle-node bifurcation in a system with a uniformly moving potential.
The nucleation time $t_{v}$ is determined by the characteristic increase of the total energy and the appearance of vortices in the density profile, as discussed in Fig. 14. 
We show the relation between $t_v$ and the frequency $\omega$ with the fixed amplitude $\epsilon = 6$ $\mu$m in Fig. 15.  
We find that $t_v$ diverges as $\omega$ approaches $\omega _{c} = 73.5$ $/s$ in Fig. 15.
This divergence means that the characteristic time becomes long when the dynamical response of the system drastically changes across $\omega _{c}$. 
However, we could not find a critical region in this work, probably because the region is too narrow to appear in our calculation. 
This region may be wide for different values of amplitude $\epsilon$. 

\section{HEATING OF THE CONDENSATE}

\begin{figure}[!t]
\begin{center}
\includegraphics[keepaspectratio, width=8cm,clip]{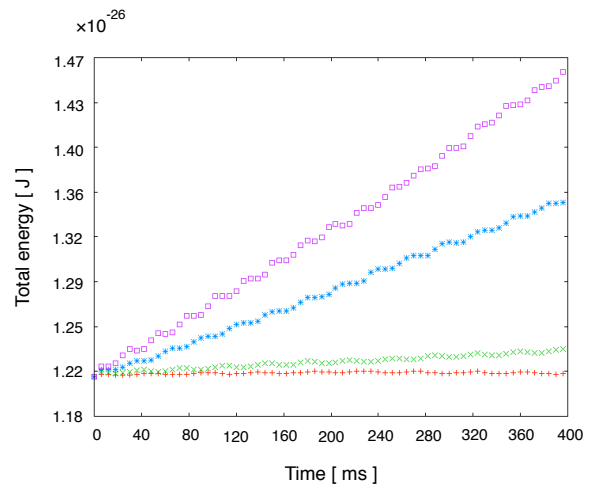}
\caption{(Color online) Increase of the total energy: The four kinds of points denote the increase of the total energy for $\omega=70$/s ($+$), $\omega=90$/s ($\times$), 
$\omega=110$/s ($*$), $\omega=130$/s ($\Box$) with $\epsilon=5$ $\mu$m. For the $+$ and $\times$ points, vortices are not nucleated, while vortices are nucleated for $*$ and $\Box$. }
\end{center}
\end{figure}

It is possible for an oscillating potential to increase the temperature of the system, and the GP model may break down. We can numerically calculate the total energy as shown in Fig. 16, qualitatively estimating the increase of the temperature by using the specific heat $C$ of the equilibrium state for an ideal Bose gas, which is expressed 
by 
\begin{equation}
C = \frac{6 k_{B}^3 T^2 \zeta (3)}{\hbar ^2 \omega _{x} \omega _{y} },
\end{equation}
where $\zeta (n)$ is the zeta function. 
In the case of $\epsilon=5 \rm \mu$m and $\omega=130$ /s in Fig. 16, the increase of the total energy is 
$2.44 \times 10^{-27}$ J for 400 ms and the specific heat is $5.40 \times 10^{-18}$ J/K, which leads to a temperature increase of $0.45$ nK. 
This is very small compared with the typical temperatures in experiments, which are well described by the GP equation. 
Therefore we conclude that the GP model is still valid against the heating in our simulations. 

Figure 16 also shows that the increase of total energy is different depending on whether the vortices are nucleated. The energy for the no vortex situation ($\omega=70$ /s and $\omega=90$ /s) slightly increases due to nucleation of phonons. However, in the case of nucleation of vortices ($\omega=110$ /s and $\omega=130$ /s), the energy greatly increases. 
Thus, there is a clear difference between the two cases, as reported by Jackson $et$ $al$.  \cite{Jackson00}. 

\section{DISCUSSION}

We have performed numerical calculations with the GP equation under the condition that the aspect ratio of the shape of the condensate is fixed. One may be concerned about the dependence of the aspect ratio
on the dynamics. However, we believe that the dependence is weak. 
The dynamics induced by an oscillating potential can be classified into two kinds. One is the dynamics near the potential, which includes the reconnection of vortices 
and nucleation of solitons near the potential. These dynamics are barely concerned with the surface, occurring with an arbitrary aspect ratio. 
The other is the dynamics of vortices and solitons related to the surface of the condensate, including the migration of vortices in Fig. 8 and the collision and collapse of solitons in Fig. 7. The migration is considered to have a weak dependence because the vortices only move along the surface. On the other hand, the collision and collapse 
can have a dependence on the aspect ratio. The direction of motion of the solitons depends on the curvature of the surface, so that the collision and collapse may not occur. 

Our numerical calculations do not contain noise in the initial state, so that all our results have mirror symmetry about the $x$ axis. 
However, our results should be consistent with experiments with noise and fluctuation breaking the symmetry of the system. 
Our results for the dynamics of vortices and solitons are dominated by their behavior near the potential, such as the reconnection of 
vortices and nucleation of solitons near the potential. These dynamics are local, so that the noise do not influence them. 
Though the noise would break the mirror symmetry of the condensate, the overall dynamics would not be qualitatively affected. 

The results in this paper are based on the two-dimensional (2D) GP equation, from which we can infer the dynamics of vortices and solitons for the three-dimensional (3D) case. 
A vortex pair and soliton with linear configuration in 2D correspond to a vortex ring and a soliton with planar configuration in the 3D case. 
Thus, the dynamics in 3D would be obtained by replacing the vortex pair and linear soliton with a vortex ring and planar soliton . An oscillating potential in 3D can create a vortex tangle in the condensate and lead to QT. 

Finally, we consider the appropriate parameters for an oscillating potential for the growth of QT. QT needs vortices and hence the regions II and III in Fig. 13 would be appropriate. 
However, region II with nucleation of solitons near the potential will not develop to QT as the number of the nucleated vortices is insufficient. 
Therefore, we suggest region II with a large amplitude compared with the size of the potential and region III as appropriate parameters for the growth of QT. 
An oscillating potential with the parameters in regions II and III may lead to granulation of the condensate. In fact, the sign appears in Fig. 12 (d), where the boundary of the 
condensate is granular.

\section{CONCLUSION}

We numerically calculated the two dimensional GP equation to investigate the dependence of the parameters on the dynamics of the condensate induced by an oscillating potential. 
We roughly obtained four kinds of dynamics. First, the oscillating potential does not nucleate vortices and solitons; only the emission of phonons occurs. Second, the potential 
nucleates vortices and reconnection of vortices near the potential in Fig. 3 occurs. Thereafter, the vortices migrate in the condensate or transform into the solitons as
shown in Sec. IV. Thirdly, the potential nucleates solitons near itself in Fig. 9, which is very different from the case in Sec. IV. After nucleation, the solitons transform into 
vortex pairs or transfer to the surface of the condensate as shown in Sec. V. Fourthly, multiple vortex pairs are nucleated by the potential, which forms the distorted condensate 
in Fig. 12. These dynamics depend on the parameters of the oscillating potential, which is represented by the phase diagram in Fig. 13.  

We minutely discuss the nucleation of vortices induced by an oscillating potential. 
The key to the nucleation is the divergence of the quantum pressure, which leads to the breakdown of Kelvin's theorem on circulation. 
Thus, nucleation requires a low density region to cause the divergence, from which it follows that the mechanism of the nucleation has two processes. 
One is the process in which the vortices are nucleated through the growth of the amplitude of phonons. The other is that the low density region inside the potential 
causing the divergence leads to nucleation of vortices without the growth of phonons. 
We propose a new expression of Eq. (9) for the critical velocity $v_c$ of the nucleation, including both the coherence length $\xi$ and the size $L$ of the potential.

We treated the nucleation of vortices by an oscillating potential as a problem of dynamical critical phenomena, and investigated the relation between the nucleation 
time and the frequency. Our numerical calculation revealed that the nucleation time diverges near the critical frequency, but we unfortunately could not confirm the power law between the time and the frequency.

\section*{ACKNOWLEDGMENT}
M. T. acknowledges the support of a Grant-in-Aid for Science Research from JSPS (Grant No. 21340104).


\begin{thebibliography}{99}
 
\bibitem{Donnelly} R. J. Donnelly, {\it Quantized Vortices in Helium II} (Cambridge University Press, Cambridge, 1991).
\bibitem{Vollhardt} D. Vollhardt and P. W\"{o}lfle, {\it The Superfluid Phases of Helium 3} (Taylor and Francis, London,1990). 
\bibitem{Cornell} M. H. Anderson, J. R. Ensher, M. R. Matthews, C. E. Wieman, and E. A. Cornell, Science \textbf{269}, 198 (1995).
\bibitem{Ketterle} K. B. Davis, M.-O. Mewes, M. R. Andrews, N. J. van Druten, D. S. Durfee, D. M. Kurn, and W. Ketterle, Phys. Rev. Lett. \textbf{75}, 3969 (1995).
\bibitem{Hulet} C. C. Bradley, C. A. Sackett, J. J. Tollett, and R. G. Hulet, Phys. Rev. Lett. \textbf{75}, 1687 (1995).
\bibitem{Tsubota02} M. Tsubota, K. Kasamatsu, and M. Ueda, Phys. Rev. A \textbf{65}, 023603 (2002). 
\bibitem{Kasamatsu03} K. Kasamatsu, M. Tsubota, and M. Ueda, Phys. Rev. A \textbf{67}, 033610 (2003). 
\bibitem{Frisch1992} T. Frisch, Y. Pomeau, and S. Rica, Phys. Rev. Lett. \textbf{69}, 1644 (1992).
\bibitem{Jackson1998} B. Jackson, J. F. McCann, and C. S. Adams, Phys. Rev. Lett. \textbf{80}, 3903 (1998).
\bibitem{Neely10} T. W. Neely, E. C. Samson, A. S. Bradley, M. J. Davis, and B. P. Anderson, Phys. Rev. Lett. \textbf{104}, 160401 (2010). 
\bibitem{Sasaki10} K. Sasaki, N. Suzuki, and H. Saito, Phys. Rev. Lett. \textbf{104}, 150404 (2010). 
\bibitem{tsuchiya} S. Tsuchiya, F. Dalfovo, and L. Pitaevskii, Phys. Rev. A \textbf{77}, 045601 (2008).
\bibitem{Anderson01} B. P. Anderson, P. C. Haljan, C. A. Regal, D. L. Feder, L. A. Collins, C. W. Clark, and E. A. Cornell, Phys. Rev. Lett. \textbf{86}, 2926 (2001).
\bibitem{Huang23} G. Huang, V. A. Makarov, and M. G. Velarde, Phys. Rev. A \textbf{67}, 023604 (2003).
\bibitem{Za} V. E. Zakharov and A. M. Rubenchik, Sov. Phys.-JETP \textbf{38}, 494 (1974).
\bibitem{Robert1982} C. A. Jones and P. H. Roberts, J. Phys. A \textbf{15}, 2599 (1982).
\bibitem{Natalia02} N. G. Berloff, Phys. Rev. B \textbf{65}, 174518 (2002).
\bibitem{fujiyama} R. Goto, S. Fujiyama, H. Yano, Y. Nago, N. Hashimoto, K. Obara, O. Ishikawa, M. Tsubota, and T. Hata, Phys. Rev. Lett. \textbf{100}, 045301 (2008).
\bibitem{s} K. Fujimoto and M. Tsubota, Phys. Rev. A \textbf{82}, 043611 (2010).
\bibitem{PLTP} W. P. Halperin and M. Tsubota, eds., {\it Progress in Low Temperature Physics} (Elsevier, Amsterdam, 2009), Vol. 16.
\bibitem{Berloff02} N. G. Berloff and B. V. Svistunov, Phys. Rev. A \textbf{66}, 013603 (2002).
\bibitem{Kobayashi07} M. Kobayashi and M. Tsubota, Phys. Rev. A \textbf{76}, 045603 (2007).
\bibitem{Henn09} E. A. L. Henn, J. A. Seman, G. Roati, K. M. F. Magalhaes, and V. S. Bagnato, Phys. Rev. Lett. \textbf{103}, 045301 (2009). 
\bibitem{Takeuchi10} H. Takeuchi, S. Ishino, and M. Tsubota, Phys. Rev. Lett. \textbf{105}, 205301 (2010). 
\bibitem{Raman99} C. Raman, M. K$\ddot{o}$hl, R. Onofrio, D. S. Durfee, C. E. Kuklewicz, Z. Hadzibabic, and W. Ketterle, Phys. Rev. Lett. \textbf{83}, 2502 (1999).
\bibitem{Onofrio00} R. Onofrio, C.Raman, J. M. Vogels, J. R. Abo-Shaeer, A. P. Chikkatur, and W. Ketterle, Phys. Rev. Lett. \textbf{85}, 2228 (2000).
\bibitem{Jackson00} B. Jackson, J. F. McCann, and C. S. Adams, Phys. Rev. A \textbf{61}, 051603 (2000). 
\bibitem{Huepe00} C. Huepe and M. Brachet, Physica D \textbf{140}, 126 (2000). 
\bibitem{pattern} M. C. Cross, and P. C. Hohenberg, Rev. Mod. Phys. \textbf{65}, 851 (1993). 
\bibitem{Ke} We can identify a point on $C$ with a time-independent parameter $s$ because of the following. A point on $C$ is specified by a one-dimensional coordinate $l(t)$ along $C$. 
This coordinate $l(t)$ depends on time since the contour $C$ changes with time. However, we can always express the coordinate $l(t)$ as a monotonously increasing function of the time-independent parameter $s$ with the domain $s_{0} \leq s \leq s_{1}$. 
\bibitem{nucleation} Divergence of the quantum pressure is a necessary condition for the nucleation of vortices, and not a sufficient condition. 
For example, a black soliton causes divergence of the quantum pressure, but the vortices cannot nucleate in a system with no noise. 
Only by performing a numerical calculation, can we find whether vortices actually nucleate. 
\bibitem{image} A. L. Fetter, J. Low Temp. Phys. \textbf{161}, 445 (2010).
\bibitem{granulation} J. A. Seman, E. A. Henn, R. F. Shiozaki, G. Roati, F. J. Poveda-Cuevas, K. M. F. Magalh$\rm \tilde{a}es$ , V. I. Yukalov, M. Tsubota, M. Kobayashi, K. Kasamatsu, V. S. Bagnato, arXiv:1007.4953. 
\bibitem{Landau} L. D. Landau and E. M. Lifschitz, \textit{Fluid Mechanics} (Pergamon, London, 1987) 2nd ed. 
\bibitem{Pines} P. Nozi$\rm \grave{e}$res and D. Pines, {\it The Theory of Quantum Liquids II} (Addison-Wesley, Redwood City, CA 1990).
\bibitem{Zwerger00} J. S. Stiessberger and W. Zwerger, Phys. Rev. A \textbf{62}, 061601 (2000). 
\bibitem{Crescimanno2000} M. Crescimanno, C. G. Koay, R. Peterson, and R. Walsworth, Phys. Rev. A \textbf{62}, 063612 (2000).
\bibitem{p} C. J. Pethick and H. Smith, {\it Bose--Einstein Condensation in Dilute Gases} (Cambridge University Press, Cambridge, 2008), Second Edition.
\bibitem{Tough} J. T. Tough, in {\it Progress in Low Temperature Physics}, edited by D. F. Brewer (North-Holland, Amsterdam, 1982), Vol. VIII, p. 133.
\bibitem{Griswold87} D. Griswold, C. P. Lorenson, and J. T. Tough, Phys. Rev. B \textbf{35}, 3149 (1987).
\bibitem{Adachi10} H. Adachi, S. Fujiyama, and M. Tsubota, Phys. Rev. B \textbf{81}, 104511 (2010).
\bibitem{Adachi} H. Adachi, private communication. 

\end{thebibliography}
\end{document}